\documentclass[a4paper,fleqn]{cas-sc}
\usepackage{soul}
\usepackage[sort&compress,numbers]{natbib}
\usepackage{lineno}
\usepackage{setspace}
\usepackage{verbatim}
\usepackage{graphics}
\usepackage{graphicx}
\usepackage{cuted}
\usepackage{lipsum}
\usepackage{multirow}

\def\tsc#1{\csdef{#1}{\textsc{\lowercase{#1}}\xspace}}
\tsc{WGM}
\tsc{QE}
\tsc{EP}
\tsc{PMS}
\tsc{BEC}
\tsc{DE}

\begin{document}

\let\WriteBookmarks\relax
\def\floatpagepagefraction{1}
\def\textpagefraction{.001}
\shorttitle{Energetic Stabilization of Janus-MoSSe/AlN Heterostructures}
\shortauthors{Ramiro Marcelo dos Santos \textit{et~al}.}

\title [mode = title]{Predicting the Energetic Stabilization of Janus-MoSSe/AlN Heterostructures: A DFT Study}
\author[1]{Ramiro M. dos Santos}
\author[1]{Marcelo L. Pereira J\'unior}
\author[1]{Luiz F. Roncaratti J\'unior}
\author[1,2]{Luiz A. Ribeiro J\'unior}
\cormark[1]
\ead{ribeirojr@unb.br}

\cortext[cor1]{Corresponding author}

\address[1]{Institute of Physics, University of Bras\'ilia, 70910-900, Bras\'ilia, Brazil}
\address[2]{PPGCIMA, Campus Planaltina, University of Bras\'{i}lia, 73345-010, Bras\'{i}lia, Brazil}

\begin{abstract}
The packing mechanisms between Janus-MoSSe and Aluminum-Nitride (AlN) sheets were systematically investigated by using Density Function Theory calculations. Results show that the stabilization (packing) energies vary from -35.5 up to -17.5 meV depending on the chemical species involved in the interface. The packing energies were obtained using the improved Lennard-Jones (ILJ) potential. The SeMoS/AlN heterostructures, when the MoS face is interacting with the AlN sheet, presented the lowest packing energies due to the sulfur's higher degree of reactivity. Importantly, the calculated bandgap values ranged within the interval 1.61--1.87 eV, which can be interesting for photovoltaic applications.                     
\end{abstract}

\begin{keywords}
Janus-MoSSe \sep Aluminum-Nitride \sep Packing Energy \sep Improved Lennard-Jones  
\end{keywords}

\maketitle
\doublespacing

\section{Introduction}

Due to the enormous growth in the global demand for energy consumption in the last decades, novel advances in renewable energy technologies have emerged recently \cite{green2002third,nozik2010semiconductor,conibeer2006silicon,ludin2014review,polman2012photonic}. In the establishment of these technologies, transition metal dichalcogenides (TMDs) have been playing an important role \cite{tsai2014monolayer, wi2014enhancement}. The most known TMD is the molybdenum disulfide (MoS$_2$), whose hexagonal monolayer (1H) phase is structurally similar to graphene \cite{dheer2020van, zan2015influence}. One of the greatest advantages of MoS$_2$ in relation to graphene is its bandgap of 1.9 eV (1.3 eV) for monolayer (multilayer) phase \cite{splendiani2010emerging, tonndorf2013photoluminescence}. These values are favorable for optical absorption when exposed to solar radiation \cite{li2015giant, steinhoff2015efficient, singh2017atomically, monolayer}. In this sense, several works have been developed aiming at designing optoelectronic devices based on MoS$_2$ \cite{yin2012single,deng2014black, li2017electronic, singh2020defect, liao2018mos}.  

TMD-based heterostructures have been both theoretical and experimentally studied \cite{PhysRevMaterials.1.044003, CHEN2014184, articlenature,ugeda2014giant,PhysRevB.92.155403,azizi2015freestanding}. Recently, Janus-MoSS2-based heterostructures emerged as promising solutions for visible-infrared photocatalysis for water splitting \cite{photocatalytic,xu2020intrinsic,idrees2020electronic} and metallic electrodes \cite{D0NR02084B}. Importantly, is was experimentally reported that a Janus-MoSSe monolayer can be obtained through breaking the out-of-plane structural symmetry of the single-layer MoS$_2$\cite{lu2017janus}. In the yielded structure, the sulfur atoms on one side of the monolayer are fully replaced by the selenium ones \cite{lu2017janus}. Yin and coworkers theoretically studied the role of the intrinsic dipole on photocatalytic water splitting for Janus-MoSSe/Nitrides heterostructures by employing Density Functional Theory (DFT) calculations \cite{YIN2019335}. Their results showed that MoSSe/XN (X=Al, Ga) configurations with a perfect match between the hexagonal rings are always more stable than other types of stacking regardless of possible atomic positions \cite{YIN2019335}. Zhao and Schwingenschl\"{o}gl investigated the van der Waals heterostructures constructed from Janus-Mosse/Germanene through first-principles calculations \cite{D0NR02084B}. The germanene layer was chosen as electrode since it is a Dirac metal with a perfect lattice match to MoSSe \cite{D0NR02084B}. Their findings revealed that an n-type Schottky contact was formed for SeMoS/Ge and a p-type Schottky contact for SMoSe/Ge \cite{D0NR02084B}. A transition from Schottky to Ohmic behavior occurs under tensile strain ($\sim$ 4\% for SeMoS/Ge, $\sim$ 8\% for SMoSe/Ge), which was explained by modifications of the interface dipole \cite{D0NR02084B}. Albeit relevant studies have been performed to propose feasible applications for Janus-MoSSe-based heterostructures, to the best of our knowledge, the fundamental properties of Janus-MoSSe/Nitrides, such as their packing stabilization, remains not fully understood.     

In the present work, we employed DFT calculations to systematically study the stabilization (packing) mechanism of Janus-MoSSe and Aluminum-Nitride (MoSSe/AlN) heterostructures. The computational protocol developed here used the improved Lennard Jones potential (ILJ) \cite{B808524B} and \textit{ab initio} Molecular Dynamics (MD) to obtain the packing energies of van der Waals heterostructures constructed from MoSSe considering both the MoS and MoSe faces interacting with the AlN sheet. The results presented here shed light on the role played by Janus-MoSSe layers in stabilizing van der Waals heterostructures based on TMDs.    

\section{Methodology}

To investigate the structural and electronic properties of MoSSe/AlN heterostructures, we employed DFT calculations as implemented in the SIESTA package \cite{Soler_2002}. Within the framework of SIESTA, we used the numerical DZP basis set to expand the system wave functions of many atoms \cite{PhysRev.140.A1133,PhysRev.136.B864,PhysRevB.64.195103}.  The exchange-correlation energies were calculated by using the Local Density Approximation (LDA), as proposed for Ceperley-Alder and Perdew-Wiang (LDA/CA and LDA/PW92, respectively) \cite{PhysRevB.28.1809}, and the Generalized Gradient Approximation (GGA), as proposed for Perdew-Burke-Ernzerhof and Perdew-Wiang (GGA/PBE, GGA/PW91, respectively) \cite{PhysRevLett.77.3865}. The relativistic pseudopotentials were parameterized within the Troullier-Martins formalism \cite{PICKETT1989115, PhysRevB.43.8861}. These approximations are required to describe the magnetic and electronic properties of materials composed of atoms with many electrons, as is the case of transition metals. A mesh cutoff of 400 Ry is chosen as a parameter for our calculations. The supercells containing MoSSe/AlN were previously converged, with a force criterion of 0.001 eV/\AA, to fit in the box that was maintained orthogonal during all optimization. To calculate the bands an MPK mesh of 9 $\times$ 9 $\times$ 3 was used \cite{PhysRevB.13.5188}.  

The following equation was used to describe the packing energy ($E_{P}$)   
\begin{equation}
\displaystyle E_{P}(r) = E_\text{AlN}^\text{MoSSe}(r) - E_\text{AlN} - E_\text{MoSSe},
\end{equation}
\noindent where $E_\text{AlN}^\text{MoSSe}(r)$ is the energy of the MoSSe/AlN system that depends on the distance between the two monolayers. $E_\text{AlN}$ and $E_\text{MoSSe}$ are the energy of the isolated AlN monolayer and the isolated MoSSe monolayer, respectively.
\noindent The minimum packing energies and equilibrium distances for all the MoSSe/AlN configurations studied here (see Figure \ref{fig:configurations}) were obtained through the fitting of the interaction energy curves using the ILJ potential  \cite{B808524B}, as presented in the following  
\begin{center}
\begin{equation}
\text{V}^{ILJ}(r) =   \epsilon \left( \frac{m}{n(r)-m} \left( \frac{r_m}{r}  \right)^{n(r)} -  \frac{n(r)}{n(r)-m}  \left( \frac{r_m}{r} \right)^{m} \right).
\label{eq:ILJ}
\end{equation} 
\end{center}
\noindent In the equation above, $\epsilon$ is the background energy of the well, and $r_m$ is the distance corresponding to $\epsilon$ (lowest $E_P$ energy in our case). Here, we assume $m=6$ to account neutral-neutral interactions \cite{B808524B}. Importantly,   
\begin{center}
\begin{equation}
n(r)= \beta +4 \left( \frac{r}{r_m} \right) ^2,
\end{equation}
\end{center}
\noindent where $\beta$ is a parameter related to the hardness of the interaction between the two systems. If $n(r)$ becomes independent of $r$, then we obtain the usual relationship for the Lennard-Jones potential \cite{B808524B}. Importantly, \textit{ab initio} MD simulations were also performed, using the SIESTA code, to study the thermodynamical stability of the MoSSe/AlN heterostructures. These simulations were performed using an NPT ensemble with an initial/target temperature set to 300 K and a time-step of 3 fs.

\section{Results}

\begin{figure}[pos=t]
\centering
\includegraphics[width=1.0\linewidth]{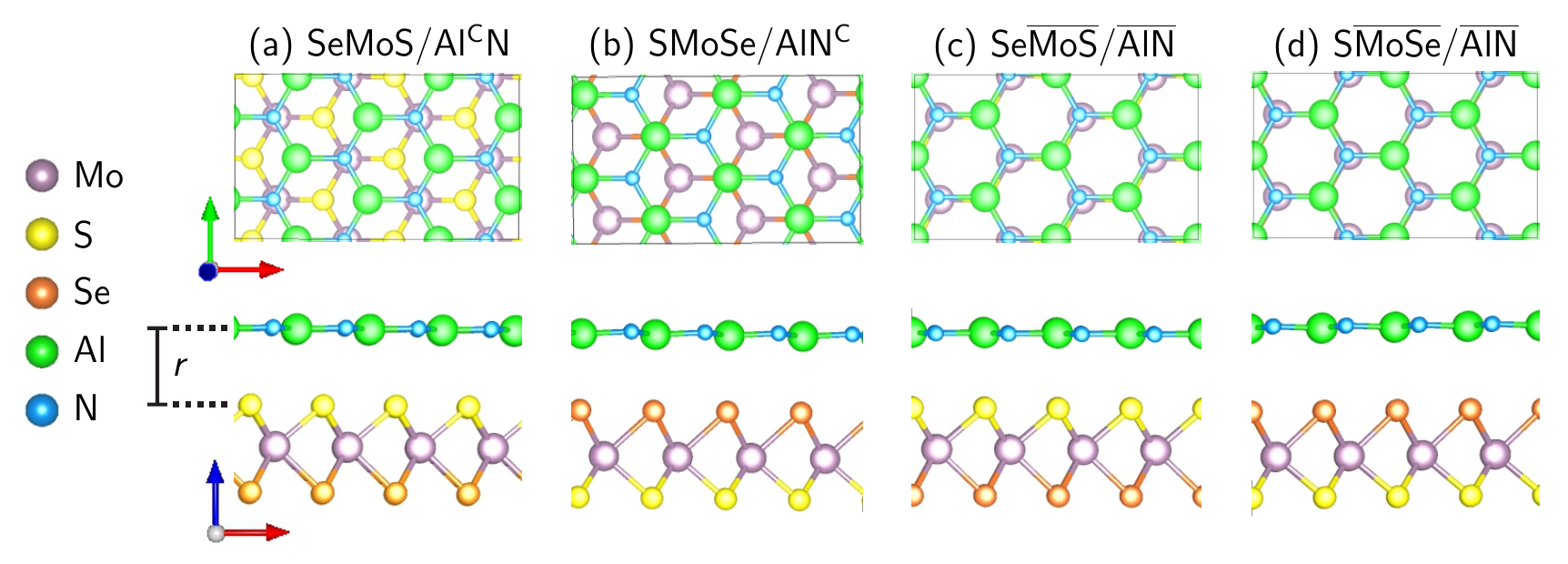}
\caption{Schematic representation of the four different cases considered in our simulations: (a) SeMoS/Al$^{\text{C}}$N (S-AlN interface), (b) SMoSe/AlN$^{\text{C}}$ (Se-AlN interface), (c) Se$\overline{\text{MoS}}$/$\overline{\text{AlN}}$ (S-AlN interface), and (d) S$\overline{\text{MoSe}}$/$\overline{\text{AlN}}$ (Se-AlN interface), where N$^{\text{C}}$ (Al$^{\text{C}}$) and $\overline{\text{AlN}}$ denote a configuration in which the nitrogen (aluminium) is placed in the center of the TMD hexagons and the nitrogen is placed vertically aligned with the molybdenum atoms, respectively.}
\label{fig:configurations}
\end{figure}

We begging our discussions by presenting the different MoSSe/AlN systems investigated here as well as the protocol used to obtain their most stable packing configurations. In Figure \ref{fig:configurations} we illustrate the four different cases considered in our simulations: \ref{fig:configurations}(a) SeMoS/Al$^{\text{C}}$N (S-AlN interface), \ref{fig:configurations}(b) SMoSe/AlN$^{\text{C}}$ (Se-AlN interface), \ref{fig:configurations}(c) Se$\overline{\text{MoS}}$/$\overline{\text{AlN}}$ (S-AlN interface), and \ref{fig:configurations}(d) S$\overline{\text{MoSe}}$/$\overline{\text{AlN}}$ (Se-AlN interface), where N$^{\text{C}}$ (Al$^{\text{C}}$) and $\overline{\text{N}}$ denote a configuration in which the nitrogen (aluminum) atoms are localized in the center of the TMD hexagons and the nitrogen atoms are vertically aligned with the molybdenum ones, respectively. These model supercells have the following dimensions: \ref{fig:configurations}(a) 10.87 $\times$ 6.30 $\times$ 49.41 \r{A}, \ref{fig:configurations}(b) 10.04 $\times$ 6.40 $\times$ 38.86 \r{A}, \ref{fig:configurations}(c) 10.86 $\times$ 6.29 $\times$ 49.51 \r{A}, and \ref{fig:configurations}(d) 10.89 $\times$ 6.31 $\times$ 39.9 \r{A}. Importantly, the bond-length values obtained here for the optimized structures (see Table \ref{tab:bonds}) are in good agreement with the ones reported in literature \cite{YIN2019335}. In our computational protocol to predict the most stable packing configuration (with lowest stabilization energy), we performed a systematic variation of the $r$ distance (see Figure \ref{fig:configurations}) between the TMD and AlN surfaces from 1.9 up to 5.6 \r{A}. The composite systems were initially optimized to adjust cell lengths and the TMD and AlN planes were separated from 3.5 \r{A}. For each $r$ distance, a single-point calculation was performed to obtain the interaction energy. Importantly, the four cases studied here (Figure \ref{fig:configurations}) are solutions of the geometry optimization procedure when the MoSSe and AlN sheets were positioned with/without matching their hexagonal rings. Moreover, \textit{ab initio} MD simulations were also performed to verify the dynamical stability of the interfaces with lowest packing energies.

\begin{table}[pos=t] 
\centering
\begin{tabular}{c c c c c }
\hline                                  
                      Bond Type &SeMoS/Al$^{\text{C}}$N  &SMoSe/AlN$^{\text{C}}$  & Se$\overline{\text{MoS}}$/$\overline{\text{AlN}}$ & S$\overline{\text{MoSe}}$/$\overline{\text{AlN}}$         \\
\hline                      
 Mo-S  (\AA)     &  2.39           &    2.42           &     2.39      & 2.39      \\
 Mo-Se (\AA)     &  2.50           &    2.51           &     2.50      & 2.48       \\
 S-Se  (\AA)     & 3.25            &    3.26           &     3.26      & 3.21       \\
Al-N   (\AA)     & 1.82            &    1.85           &     1.81      & 1.83       \\     
\hline
\end{tabular}
\caption{Bond-length values obtained here for the optimized MoSSe and AlN structures. Importantly, the values presented here are in good agreement with related ones reported in literature \cite{YIN2019335}.}	
\label{tab:bonds}
\end{table}

\begin{figure}[pos=t]
\centering
\includegraphics[width=1.0\linewidth]{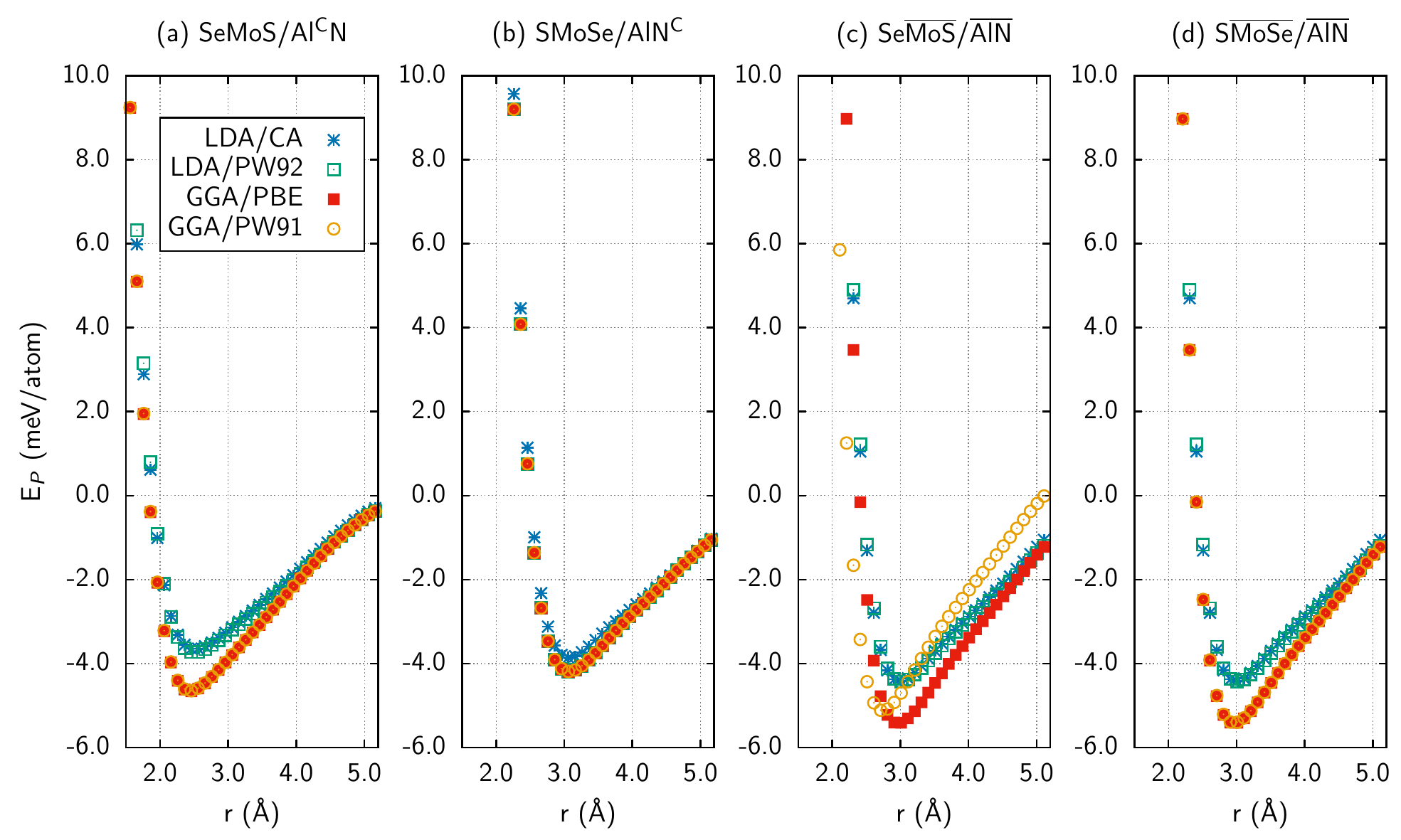}
\caption{Stabilization energy curves obtained using the set of DFT functionals and potentials described in the previous section.}
\label{fig:packcurves}
\end{figure}

The stabilization energy curves are shown in Figure \ref{fig:packcurves}. These curves were obtained using the set of DFT functionals and potentials described in the previous section. In Figure \ref{fig:packcurves}, one can note that the lowest $E_P$ energy for the SeMoS/Al$^{\text{C}}$N, SMoSe/AlN$^{\text{C}}$, Se$\overline{\text{MoS}}$/$\overline{\text{AlN}}$, and S$\overline{\text{MoSe}}$/$\overline{\text{AlN}}$ cases is -4.66 meV (LDA/CA), -4.20 meV (LDA/CA), -5.41 meV (LDA/CA), and -5.50 meV (LDA/PW92), respectively. As expected, the cases with the lowest packing energies are the ones in which the perfect match among the hexagons of both structures takes place (Se$\overline{\text{MoS}}$/$\overline{\text{AlN}}$ and S$\overline{\text{MoSe}}$/$\overline{\text{AlN}}$ cases). The cases where the sulfur atoms are interacting with the AlN surface (SeMoS/Al$^{\text{C}}$N and Se$\overline{\text{MoS}}$/$\overline{\text{AlN}}$) present the lowest packing energies when contrasted to the selenium ones due to the higher degree of reactivity presented by silicon. Among all the heterostructures, the case with perpendicular configuration between sulfur-aluminum (Se$\overline{\text{MoS}}$/$\overline{\text{AlN}}$) presents the higher packing energy than the related one with unpaired hexagons (SeMoS/Al$^{\text{C}}$N) due to the lower sulfur-aluminum distances.  

\begin{figure}[pos=t]
\centering
\includegraphics[width=1.0\linewidth]{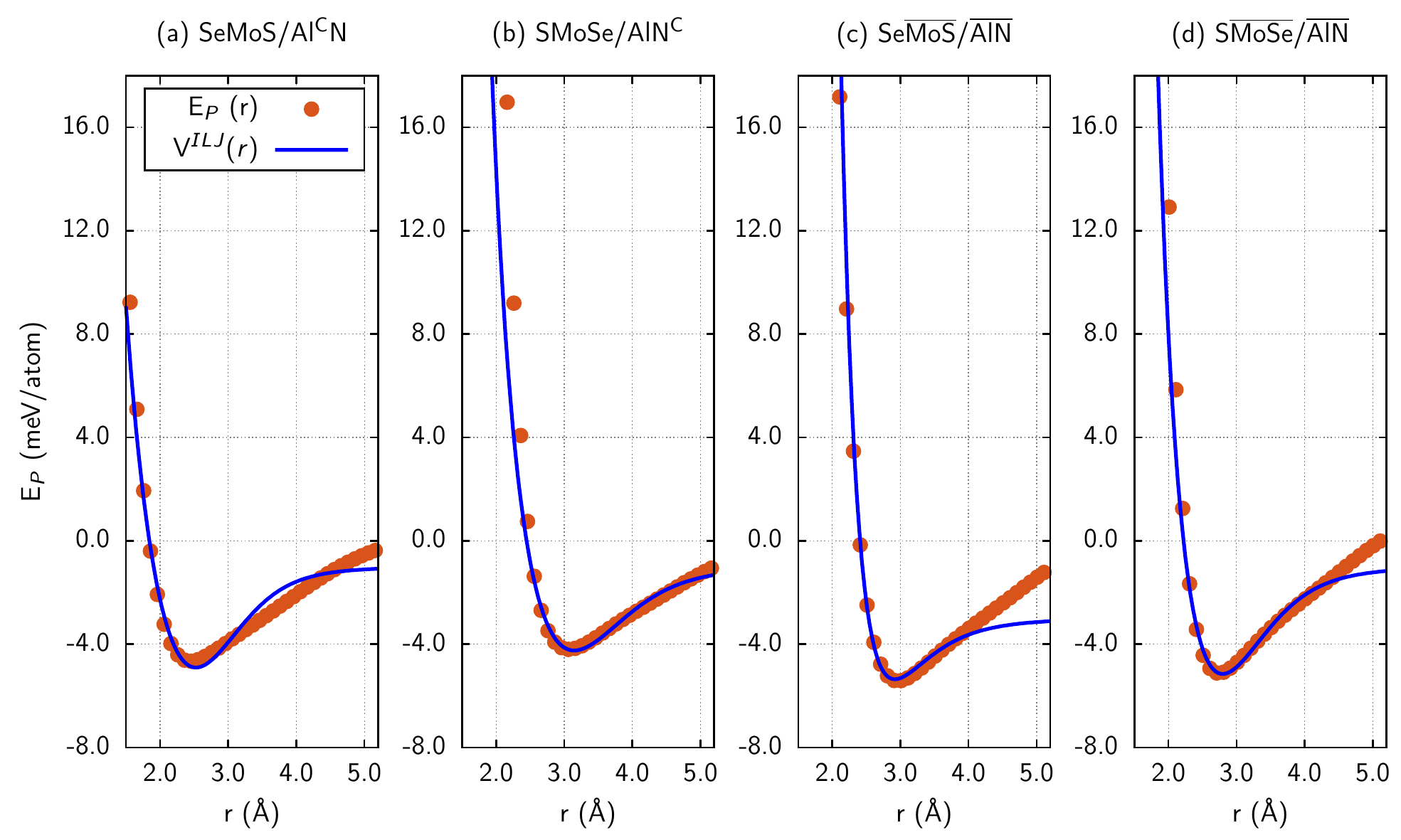}
\caption{Fitting for the stabilization energy curves for the cases with the lowest packing energies in Figure \ref{fig:packcurves}, using the ILJ potential (see Equation \ref{eq:ILJ}) \cite{B808524B}.}
\label{fig:fitting}
\end{figure}

Figure \ref{fig:fitting} shows the fitting for the stabilization energy curves of lowest packing energies in Figure \ref{fig:packcurves}, using the ILJ potential (see Equation \ref{eq:ILJ}) \cite{B808524B}. In Figure \ref{fig:fitting}, one can note that the ILJ potential accurately describes the curves wheels, predicting packing energies close to the ones obtained in the DFT calculations (see Table \ref{tab:ILJ}). 

\begin{table}[pos=t] 
\centering
\begin{tabular}{c c c c c}
\hline                                  
                       &SeMoS/AlN$^{\text{C}}$  &SMoSe/AlN$^{\text{C}}$  & Se$\overline{\text{MoS}}$/$\overline{\text{AlN}}$ & S$\overline{\text{MoSe}}$/$\overline{\text{AlN}}$         \\
\hline                                                                                                                                          
  $\epsilon$ (meV) &    4.90 ($\pm 0.10$) &  5.35 ($\pm 0.05$) & 5.15 ($\pm 0.06$)                  & 2.36 ($\pm 0.11$)                             \\
 r$_m$ (\r{A})    &    2.53 ($\pm 0.02$) &  3.15 ($\pm 0.01$) & 2.92 ($\pm 0.03$)                  & 2.80 ($\pm 0.01$                             \\
\hline
\end{tabular}
\caption{ILJ fitting parameters $\epsilon$ and $r_m$ (equilibrium distance and energy, respectively) for the stabilization energy curves for the cases with the lowest packing energies in Figure \ref{fig:packcurves}.}
\label{tab:ILJ}
\end{table}

\begin{figure}[pos=t]
\centering
\includegraphics[width=1.0\linewidth]{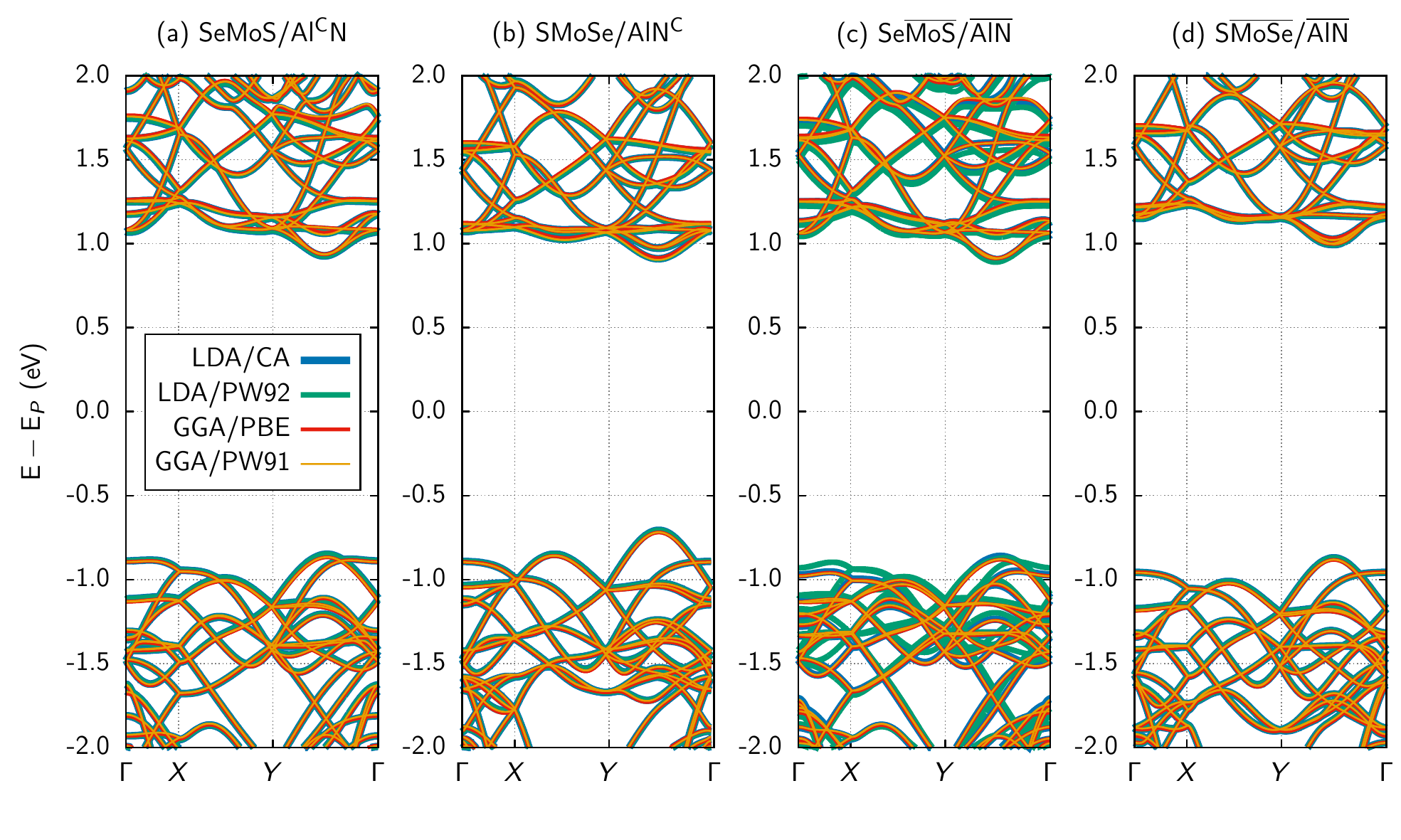}
\caption{Band structure profiles for the cases presented in Figure \ref{fig:configurations} that were obtained using the different DFT methodologies described in the previous section.}
\label{fig:bands}
\end{figure}

To verify the suitability of the DFT methodologies employed here, we contrast the band structure profiles for the cases presented in Figure \ref{fig:configurations}, that were calculated using the methods described in the previous section. The band structures were obtained by considering the cases with the lowest packing energies, which were obtained through the fitting procedure presented in Figure \ref{fig:fitting}. One can note that all the DFT methodologies present very similar band structure profiles, as illustrated in Figure \ref{fig:bands}. The bandgap values obtained using each one of them are presented in Table \ref{tab:gap}. They lie within the range of the visible spectrum and present a direct character between the Y and $\Gamma$ points, making possible the photon absorption conserving electron momentum. Importantly, the calculated bandgap values ($E_{gap}$) ranged within the interval 1.61--1.87 eV, which can be interesting for photovoltaic applications.  

\begin{table}[pos=t] 
\centering
\begin{tabular}{c c c c c }
\hline                                  
 DFT Method     &SeMoS/Al$^{\text{C}}$N  &SMoSe/AlN$^{\text{C}}$ & Se$\overline{\text{MoS}}$/$\overline{\text{AlN}}$ & S$\overline{\text{MoSe}}$/$\overline{\text{AlN}}$         \\
\hline                                                                                                                                        
 LDA/CA - E$_{gap}$ (eV)   &   1.78    &    1.62        &  1.76                   &          1.86                       \\
 LDA/PW92 - E$_{gap}$ (eV) &   1.78    &    1.61        &  1.76                   &          1.87                       \\ 
 GGA/PBE - E$_{gap}$ (eV)  &  1.80     &    1.62        &  1.79                   &          1.87                       \\
 GGA/PW91 - E$_{gap}$ (eV) &  1.80     &    1.61        &  1.80                   &          1.87                       \\
\hline
\end{tabular}
\caption{Bandgap values obtained for all the DFT methodologies used here. These values are related to the cases with the lowest packing energies, obtained through the fitting procedure presented in Figure \ref{fig:fitting}.}
\label{tab:gap}
\end{table}
 
\begin{figure}[pos=ht]
\centering
\includegraphics[width=1.0\linewidth]{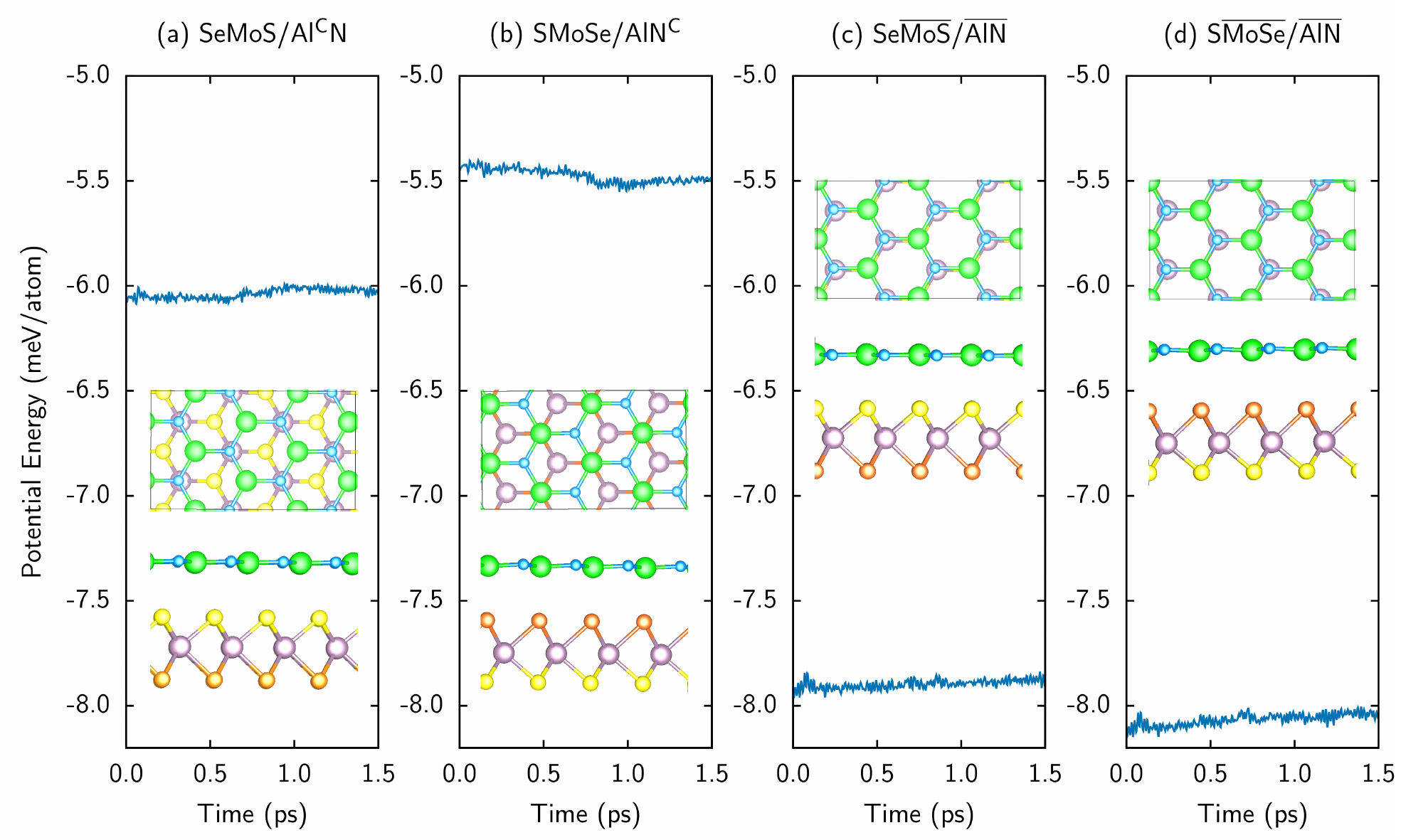}
\caption{Potential energy as a function of the time obtained in the \textit{ab initio} MD simulations that considered as input the structures with the lowest packing energies (also keeping the related DFT method).}
\label{fig:md}
\end{figure}

To validate the thermodynamical stability of the MoSSe/AlN heterostructures studied here, we performed \textit{ab initio} MD simulations, in which we considered as input the structures with the lowest packing energies (also keeping the related DFT method). In this sense, Figure \ref{fig:md} shows the time evolution of the total potential energy for each system. One can note that the potential energy remains constant during the time. As expected, the systems with lowest potential energies are the ones with a perfect match among the hexagonal rings of both structures (Se$\overline{\text{MoS}}$/$\overline{\text{AlN}}$ and S$\overline{\text{MoSe}}$/$\overline{\text{AlN}}$ cases), since they present the lowest packing energies (see Figures \ref{fig:packcurves} and \ref{fig:fitting}). The \textit{ab initio} MD results suggest that the interaction between MoSSe and AlN layers can yield stable heterostructures. It is worthwhile to stress that these results are in good agreement with the ones reported in literature \cite{YIN2019335}, which showed that MoSSe/XN (X=Al,Ga) configurations with a perfect match between the hexagonal rings are always more stable than other types of stacking regardless of possible atomic positions.

\section{Conclusions}

In summary, we used DFT calculations to theoretically investigate the packing mechanism of MoSSe/AlN heterostructures. The computational protocol employed here was based on the ILJ potential and \textit{ab inito} MD simulations to predict the packing energies of van der Waals heterostructures constructed from MoSSe, where both the MoS and MoSe faces interacted with the AlN sheet. The results revealed that the stabilization (packing) energies vary from -35.5 up to -17.5 meV depending on the chemical species involved in the interface. The lowest packing energy for the SeMoS/Al$^{\text{C}}$N, SMoSe/AlN$^{\text{C}}$, Se$\overline{\text{MoS}}$/$\overline{\text{AlN}}$, and S$\overline{\text{MoSe}}$/$\overline{\text{AlN}}$ cases is -4.66 meV, -4.20 meV, -5.41 meV, and -5.50 meV, respectively. The cases with the lowest packing energy are the ones in which the perfect match among the hexagons of both structures takes place (Se$\overline{\text{MoS}}$/$\overline{\text{AlN}}$ and S$\overline{\text{MoSe}}$/$\overline{\text{AlN}}$ cases). The cases where the sulfur atoms are interacting with the AlN surface (SeMoS/Al$^{\text{C}}$N and Se$\overline{\text{MoS}}$/$\overline{\text{AlN}}$) present the lowest packing energy when contrasted to the selenium ones due to the higher degree of reactivity presented by silicon. The stabilization energy curves were fitted by using the ILJ potential \cite{B808524B}. It was obtained that the ILJ potential can accurately describe the curves wheels, predicting packing energies close to the ones obtained in the DFT calculations. Importantly, all the DFT methodologies used here presented similar band structure profiles. The bandgap values are in the range of 1.61--1.87 eV (visible spectrum) with a direct character. These band structure features can be interesting for photovoltaic applications. Moreover, \textit{ab initio} MD simulations were performed to validate the thermodynamical stability of the MoSSe/AlN heterostructures studied here. In these simulations, the total potential energies remained constant during the time. Particularly, these results suggest that the interaction between MoSSe and AlN layers can yield stable heterostructures.

\section*{Acknowledgments}
The authors gratefully acknowledge the financial support from Brazilian Research Councils CNPq, CAPES, and FAPDF and CENAPAD-SP for providing the computational facilities. L.A.R.J. gratefully acknowledges respectively, the financial support from FAP-DF  and CNPq grants 00193.0000248/2019-32 and 302236/2018-0.

\printcredits
\bibliographystyle{unsrt}
\bibliography{cas-refs}

\end{document}